\documentclass{article}
\usepackage{eurosym}
\usepackage{amssymb}
\usepackage{amsmath}
\usepackage{cite}
\usepackage{graphicx}
\usepackage{epsf}
\usepackage{lscape}

\begin{document}

\title{Similarity transformations and linearization for a family of
dispersionless integrable PDEs}
\author{Andronikos Paliathanasis\thanks{%
Email: anpaliat@phys.uoa.gr} \\
{\ \textit{Institute of Systems Science, Durban University of Technology }}\\
{\ \textit{PO Box 1334, Durban 4000, Republic of South Africa}\ } \\
{\textit{Instituto de Ciencias F\'{\i}sicas y Matem\'{a}ticas,}}\\
{\ \textit{Universidad Austral de Chile, Valdivia, Chile}\ }}
\maketitle

\begin{abstract}
We apply the theory of Lie point symmetries for the study of a family of
partial differential equations which are integrable by the hyperbolic
reductions method and are reduced to members of the Painlev\'{e}
transcendents. The main results of this study is that from the application
of the similarity transformations provided by the Lie point symmetries all
the members of the family of the partial differential equations are reduced
to second-order differential equations which are maximal symmetric and can
be linearized.

\bigskip

\end{abstract}

\section{Introduction}

\label{sec1}

In \cite{hd01}, Ferapontov et al. classified the partial differential
equations of the form
\begin{equation}
\left( A\left( u\right) \right) _{xx}+\left( B\left( u\right) \right)
_{yy}+\left( C\left( u\right) \right) _{yy}+2\left( \left( P\left( u\right)
\right) _{xy}+\left( Q\left( u\right) \right) _{xt}+\left( P\left( u\right)
\right) _{yt}\right) =0,  \label{ee.00}
\end{equation}%
which are integrable under the method of hydrodynamic reductions \cite{hd02}
and can be reduced into an Painlev\'{e} equation \cite{ince}. There are five
partial differential equations of the form (\ref{ee.00}) which are
integrable by the hydrodynamic reductions method \cite{hd01}
\begin{equation}
\mathcal{H}_{A}\equiv u_{xx}+u_{yy}-\left( \ln \left( e^{u}-1\right) \right)
_{yy}-\left( \ln \left( e^{u}-1\right) \right) _{tt}=0,  \label{ee.01}
\end{equation}%
\begin{equation}
\mathcal{H}_{B}\equiv u_{xx}+u_{yy}-\left( e^{u}\right) _{tt}=0,
\label{ee.02}
\end{equation}%
\begin{equation}
\mathcal{H}_{C}\equiv \left( e^{u}-u\right) _{xx}+2u_{xy}+\left(
e^{u}\right) _{tt}=0,  \label{ee.03}
\end{equation}%
\begin{equation}
\mathcal{H}_{D}\equiv u_{xt}-\left( uu_{x}\right) _{x}-u_{yy}=0,
\label{ee.04}
\end{equation}%
\begin{equation}
\mathcal{H}_{E}=\left( u^{2}\right) _{xx}+u_{yy}+2u_{xt}=0.  \label{ee.05}
\end{equation}

Equations (\ref{ee.02}) and (\ref{ee.04}) are the Boyer-Finley \cite{bf0,bf1}
and dispersionless Kadomtsev-Petviashvili \cite{dkp1} equations
respectively. For these two equations it is known that they are reduced into
the Painlev\'{e} transcendents by applying \ the central quadric ansatz. The
hydrodynamic reductions method was found to provide more general solutions.
Indeed, by studying the dispersionless Kadomtsev-Petviashvili \ with the
hydrodynamic reductions and the central quadric ansatz, it was found that
the solutions coming from the later method form a subclass of two-phase
solutions provided by the hydrodynamic reductions approach \cite{hd01}.

As far as the reduction of equations (\ref{ee.01})-(\ref{ee.05}) into a
Painlev\'{e} equation, is concerned, it was found that equation $\mathcal{H}%
_{A}$ reduces to the Painlev\'{e} $P_{VI}$ equation, the Boyer-Finley
equation $\mathcal{H}_{B}$ reduces into the $P_{V}$ equation reducible to $%
P_{III}$. Moreover, $\mathcal{H}_{C}$ reduces to $P_{V}$, the dispersionless
Kadomtsev-Petviashvili is related with the $P_{II}$ with a reduction to $%
P_{I}$, while the fifth equation $\mathcal{H}_{E}$ is reduced to $P_{IV}$~%
\cite{hd01}. For extensions of the results of \cite{hd01} and a connection
of the hydrodynamic reductions method with the conformal structure of
Einstein-Weyl geometry we refer the reader to \cite{hd03}.

In this work we apply the Lie symmetry analysis \cite{ovsi,ibra,Bluman,olver}
in order to investigate the algebraic properties and the similarity
transformations for the five partial differential equations (\ref{ee.01})-(%
\ref{ee.05}). The method of Lie symmetries of differential equations
established by Sophus Lie at the end of the 19th century, and provides a
systematic approach for the study and determination of solutions and
conservation laws for nonlinear differential equations.

The novelty of symmetry analysis is that invariant functions can be
determined for a given differential equation. From the invariant functions
we can define similarity transformations which are necessary to simplify the
differential equation. The similarity transformations are used to reduce the
given differential equation into an equivalent equation with less dynamical
variables. In the case of partial differential equations the independent
variables are reduces, while in the case of ordinary differential equations,
the order of the equation, that is, the dependent variables, are reduced.
There is a plethora of applications in the literature on the symmetry
analysis of various dynamical systems. The method of symmetry analysis is
applied in various systems of fluid dynamics in the studies \cite%
{l6,l11,l12,l15,mel1,mel2,mel4,tw9,mh1,mh6,km2,km3,l1}. The Burgers-heat
system is investigated by applying the symmetry analysis in the studies \cite%
{l2,l3}. A recent application of the Lie symmetry approach on
time-fractional systems is presented in \cite{tw11}. However, Lie symmetries
are very useful and for the study of ordinary differential equations. Some
studies on the symmetry analysis on the geodesic equations in curved spaces
are presented in \cite{l5,l10,gr4,gr5}. Finally, in \cite{gr6} a discussion
is given on the novelty on the application of the Lie symmetry analysis in
gravitational physics and cosmology.

Another important application of the Lie symmetry approach is the
classification scheme of differential equations according to the admitted
group of symmetries, and to the construction of equivalent transformation
which transform a given differential equation into another differential
equation of the same order, when the admitted\ Lie symmetries form the same
Lie algebra \cite{syc1,syc2,syc3}. Recently, in \cite{syc4} the authors
investigated which of the six ordinary differential equations of the Painlev%
\'{e} transcendents admit nontrivial Lie point symmetries. It was found that
equations $P_{III}$, $P_{V}$ and $P_{VI}$ have nontrivial symmetries for
special values of the free parameters. On the other hand, in \cite{nuc1} the
method of Jacobi last multiplier applied in order to determine generalized
symmetries for a particular case of the $P_{XIV}$ equation. By using
generalized-hidden symmetries, the linearization of the Painleve-Ince
equation was proved in \cite{nuc2}. The plan of the paper in as follows.

In Section \ref{sec2} we present the basic properties and definitions for
the Lie symmetry analysis of differential equations. In\ Sections \ref{sec3}-%
\ref{sec7} we determine the Lie point symmetries for the five equations of
our analysis. We determine the commutators and the Adjoint representation
such that to derive, when it is feasible, the one-dimensional optimal
system. Equation $\mathcal{H}_{A}$ admits a finite dimensional Lie algebra
of dimension four, in particular the $A_{4,5}$ in the Patera-Winternitz
classification scheme \cite{pat}. However, the rest of the equations admit
infinity Lie point symmetries. We were able to define four-dimensional Lie
subalgebras. The main observation of this work is that the application of
the Lie invariants define similarity transformation where the partial
differential equations reduce to maximal symmetric second-order equations.\
That is an important result because we are able to investigate the
integrability properties of equations (\ref{ee.01})-(\ref{ee.05}) by using
the symmetry analysis. In Section \ref{sec8} we summarize our results and we
draw our conclusions. The main result of this analysis is given in a
proposition where we show that equations (\ref{ee.01})-(\ref{ee.05}) can be
linearized with the application of Lie invariants.

\section{Preliminaries}

\label{sec2}

Assume the infinitesimal one-parameter point transformation%
\begin{eqnarray*}
t^{\prime } &=&t+\varepsilon \xi ^{t}\left( t,x,y,u\right) , \\
x^{\prime } &=&x+\varepsilon \xi ^{x}\left( t,x,y,u\right) , \\
y^{\prime } &=&y+\varepsilon \xi ^{y}\left( t,x,y,u\right) , \\
u^{\prime } &=&u+\varepsilon \eta \left( t,x,y,u\right) ,
\end{eqnarray*}%
\qquad where $\varepsilon $ is the infinitesimal parameter, $\varepsilon
^{2}\rightarrow 0$, and infinitesimal generator%
\begin{equation*}
X=\xi ^{t}\left( t,x,y,u\right) \partial _{t}+\xi ^{x}\left( t,x,y,u\right)
\partial _{x}+\xi ^{y}\left( t,x,y,u\right) \partial _{y}+\eta \left(
t,x,y,u\right) \partial _{u}.
\end{equation*}

We define the second extension $X^{\left[ 2\right] }$ of $X$ in the jet
space $\left\{
t,x,y,u,u_{t},u_{x},u_{y},u_{tt},u_{xx},u_{yy},u_{tx},u_{xy}\right\} $ as
follows
\begin{equation*}
X^{\left[ 2\right] }=\xi ^{\mu }\partial _{\mu }+\eta ^{A}\partial _{A}+\eta
^{A\left[ 1\right] }\partial _{A_{,}\mu }+\eta ^{A\left[ 2\right] }\partial
_{A_{,\mu \nu }}.
\end{equation*}%
in which $\eta ^{A\left[ 1\right] },~\eta ^{A\left[ 2\right] }$ are defined
as
\begin{equation*}
\eta ^{\left[ n\right] }=D_{\mu }\eta ^{\left[ n-1\right] }-u_{\mu _{1}\mu
_{2},...,\mu _{n-1}}D_{\mu }\left( \xi ^{\mu }\right) ,
\end{equation*}%
where $\mu =\left( t,x,y\right) $.

By definition a partial differential equation $\mathcal{H}=\mathcal{H}\left(
t,x,y,u,u_{t},u_{x},u_{y},u_{tt},u_{xx},u_{yy},u_{tx},u_{xy}\right) $ is
invariant under the action of the one-parameter point transformation with
infinitesimal generator the vector field $X$ if and only if there exist a
function $\lambda $ such that \cite{ovsi,ibra,Bluman,olver}%
\begin{equation*}
\mathcal{L}_{X^{\left[ 2\right] }}\left( \mathcal{H}\right) =\lambda
\mathcal{H}\mathbf{,~}
\end{equation*}%
in which $L_{X^{\left[ 2\right] }}$ is the Lie derivative with respect to
the vector field $X^{\left[ 2\right] }.$

Lie symmetries are mainly applied for the construction of similarity
transformations. The latter are necessary in order to simplify a given
differential equation by mean of reduction. The exact and analytic solutions
which are determined by the application of the Lie symmetries are known as
similarity solutions.

In order to perform a complete derivation of all the possible similarity
solutions we should find the admitted one-dimensional optimal system.
Consider the $n$-dimensional Lie algebra $G_{n}$\ with elements $%
X_{1},~X_{2},~...,~X_{n}$\ admitted by the differential equation $\mathcal{H}
$.

The vector fields \cite{ovsi,ibra,Bluman,olver}
\begin{equation*}
Z=\sum\limits_{i=1}^{n}a_{i}X_{i}~,~W=\sum\limits_{i=1}^{n}b_{i}X_{i}~,~%
\text{\ }a_{i},~b_{i}\text{ are constants.}
\end{equation*}%
are equivalent if and only if $\mathbf{W}=Ad\left( \exp \left( \varepsilon
_{i}X_{i}\right) \right) \mathbf{Z~}$or~$\mathbf{W}=c\mathbf{Z}~\ $where$~c$
is a constant.

Operator $Ad\left( \exp \left( \varepsilon X_{i}\right) \right)
X_{j}=X_{j}-\varepsilon \left[ X_{i},X_{j}\right] +\frac{1}{2}\varepsilon
^{2}\left[ X_{i},\left[ X_{i},X_{j}\right] \right] +...~$is known as the the
adjoint representation. The derive all the independent similarity
transformations for a given differential equation the adjoint representation
of the admitted Lie algebra should be determined. This lead to the
construction of the one-dimnesional optimal system.

\section{Lie symmetry analysis for equation $\mathcal{H}_{A}$}

\label{sec3}

The first equation of our analysis, equation $\mathcal{H}_{A},$ admits the
Lie point symmetries
\begin{equation*}
X_{1}^{A}=\partial _{t}~,~X_{2}^{A}=\partial _{x}~,~X_{3}^{A}=\partial
_{y}~,~X_{4}^{A}=t\partial _{t}+x\partial _{x}+y\partial _{y}
\end{equation*}%
with commutators and Adjoint representation as presented in Tables \ref{ha1}
and \ref{ha2}. The admitted Lie algebra is the $A_{4,5}$. The
one-dimensional system for the finite Lie algebra is consisted by the
one-dimensional Lie algebras: $\left\{ X_{1}^{A}\right\} $, $\left\{
X_{2}^{A}\right\} $, $\left\{ X_{3}^{A}\right\} $, $\left\{
X_{4}^{A}\right\} $, $\left\{ X_{1}^{A}+\alpha X_{2}^{A}\right\} $, $\left\{
X_{1}^{A}+\alpha X_{3}^{A}\right\} $,~$\left\{ X_{2}^{A}+\alpha
X_{3}^{A}\right\} $ and $\left\{ X_{1}^{A}+\alpha X_{2}^{A}+\beta
X_{3}^{A}\right\} $.

\begin{table}[tbp] \centering%
\caption{Commutators for the Lie point symmetries of equation
(\ref{ee.01}).}%
\begin{tabular}{ccccc}
\hline\hline
$\left[ X_{I},X_{J}\right] $ & \textbf{$X^{A}$}$_{1}$ & \textbf{$X^{A}$}$%
_{2} $ & \textbf{$X^{A}$}$_{3}$ & \textbf{$X^{A}$}$_{4}$ \\ \hline
\textbf{$X^{A}$}$_{1}$ & $0$ & $0$ & $0$ & $X_{1}^{A}$ \\
\textbf{$X^{A}$}$_{2}$ & $0$ & $0$ & $0$ & $X_{2}^{A}$ \\
\textbf{$X^{A}$}$_{3}$ & $0$ & $0$ & $0$ & $X_{3}^{A}$ \\
\textbf{$X^{A}$}$_{4}$ & $-X_{1}^{A}$ & $-X_{2}^{A}$ & $-X_{3}^{A}$ & $0$ \\
\hline\hline
\end{tabular}%
\label{ha1}%
\end{table}%

\begin{table}[tbp] \centering%
\caption{Adjoint representation for the Lie point symmetries of equation
(\ref{ee.01}).}%
\begin{tabular}{ccccc}
\hline\hline
$Ad\left( e^{\left( \varepsilon \mathbf{X}_{i}\right) }\right) \mathbf{X}%
_{j} $ & \textbf{$X^{A}$}$_{1}$ & \textbf{$X^{A}$}$_{2}$ & \textbf{$X^{A}$}$%
_{3}$ & \textbf{$X^{A}$}$_{4}$ \\ \hline
\textbf{$X^{A}$}$_{1}$ & $X_{1}^{A}$ & $X_{2}^{A}$ & $X_{3}^{A}$ & $%
X_{4}^{A}-\varepsilon X_{1}^{A}$ \\
\textbf{$X^{A}$}$_{2}$ & $X_{1}^{A}$ & $X_{2}^{A}$ & $X_{3}^{A}$ & $%
X_{4}^{A}-\varepsilon X_{2}^{A}$ \\
\textbf{$X^{A}$}$_{3}$ & $X_{1}^{A}$ & $X_{2}^{A}$ & $X_{3}^{A}$ & $%
X_{4}^{A}-\varepsilon X_{3}^{A}$ \\
\textbf{$X^{A}$}$_{4}$ & $e^{\varepsilon }X_{1}^{A}$ & $e^{\varepsilon
}X_{2}^{A}$ & $e^{\varepsilon }X_{3}^{A}$ & $X_{4}^{A}$ \\ \hline\hline
\end{tabular}%
\label{ha2}%
\end{table}%

Application of $\left\{ X_{1}^{A}+\alpha X_{2}^{A}\right\} $ , $\left\{
X_{1}^{A}+\beta X_{3}^{A}\right\} $ provides the reduced equation%
\begin{equation}
\left( \beta ^{2}\left( e^{U}-1\right) ^{2}-\alpha ^{2}\left( e^{U}-1\right)
\left( 1+\beta ^{2}e^{U}\right) \right) U_{\,\sigma \sigma }+\alpha
^{2}\left( \beta ^{2}+1\right) e^{U}\left( U_{\sigma }\right) ^{2}=0,
\label{ee.06}
\end{equation}%
with $u=U\left( \sigma \right) $ and $\sigma =y-\beta t+\frac{\beta }{\alpha
}x$.

Equation (\ref{ee.06}) is a second-order ordinary differential equation of
the form $U_{\sigma \sigma }+L\left( U\right) \left( U_{\sigma }\right)
^{2}=0$, which means that is maximally symmetric. It admits eight Lie point
symmetries which form the $sl\left( 3,R\right) $ algebra. Thus, according to
the main theorem of S. Lie theorem, equation (\ref{ee.06}) can be linearized
\cite{ovsi,ibra,Bluman,olver}. Indeed the transformation which linearized
the differential equation is of the form $V=\int e^{\int L\left( U\right)
dU}dU$. \ We remark that any reduction of equation (\ref{ee.01}) with any
Lie symmetries provided by the optimal system of the Abelian Lie subalgebra $%
\left\{ X_{1}^{A},X_{2}^{A},X_{3}^{A}\right\} $ provides a similar result.

On the other hand, reduction with $\left\{ X_{4}^{A}\right\} $ provides the
partial differential equation
\begin{eqnarray}
0 &=&\left( e^{U}-1\right) \left( U_{\zeta \zeta }\left( \zeta
^{2}e^{U}+1\right) +U_{\omega \omega }\left( e^{U}\left( \omega
^{2}-1\right) +1\right) +2\omega \zeta e^{U}U_{\zeta \omega }\right)  \notag
\\
&&+e^{U}\left( \left( U_{\zeta }\zeta +U_{\omega }\omega \right)
^{2}-2\left( e^{U}-1\right) \left( U_{\zeta }\zeta +U_{\omega }\omega
\right) +U_{\zeta \zeta }^{2}\right) ,  \label{ee.07}
\end{eqnarray}%
with $u=U\left( \zeta ,\omega \right) ~$, $\zeta =\frac{y}{t}$ and $\omega =%
\frac{x}{t}$, is the similarity transformation. Equation (\ref{ee.07}) does
not possess any Lie point symmetry, thus we can not reduce further the
differential equation.

Moreover, reduction with the symmetry vectors $\left\{ X_{2}^{A}+\alpha
X_{3}^{A}\right\} $ and later with the reduced symmetry of $\left\{
X_{4}^{A}\right\} $, provides the ordinary differential equation%
\begin{equation}
\left( \alpha ^{2}\left( e^{U}-1\right) -1-e^{U}\rho ^{2}\right) U_{\rho
\rho }+\frac{e^{U}}{\left( e^{U}-1\right) }\left( \rho ^{2}+1\right) \left(
U_{\rho }\right) ^{2}-2\rho e^{U}U_{\rho }=0,  \label{ee.08}
\end{equation}%
where $u=U\left( \rho \right) $ and $\rho =\frac{y-\alpha x}{t}$.

Without loss of generality we assume $\alpha ^{2}=1$. Then equation (\ref%
{ee.08}) is written in the equivalent form%
\begin{equation*}
\frac{U_{\rho \rho }}{U_{\rho }}+\frac{e^{U}\left( \rho ^{2}+1\right) }{%
\left( e^{U}-1\right) \left( e^{U}\left( 1-\rho ^{2}\right) -2\right) }%
\left( U_{\rho }\right) -\frac{2e^{U}}{\left( e^{U}\left( 1-\rho ^{2}\right)
-2\right) }=0,
\end{equation*}%
that is
\begin{equation*}
\frac{U_{\rho \rho }}{U_{\rho }}+\frac{e^{U}}{e^{U}-1}U_{\rho }+\frac{%
U_{\rho }e^{U}\rho ^{2}+2e^{U}\rho -U_{\rho }e^{U}}{e^{U}\left( \rho
^{2}-1\right) +2}=0.
\end{equation*}

Hence, we can write easily the later equation as follows
\begin{equation*}
\frac{d}{d\rho }\left( \ln \left( U_{\rho }\right) -\ln \left(
e^{U}-1\right) +\ln \left( e^{U}\left( \rho ^{2}-1\right) +2\right) \right)
=0.
\end{equation*}%
Therefore, the conservation law is%
\begin{equation}
\frac{U_{\rho }\left( e^{U}\left( \rho ^{2}-1\right) +2\right) }{e^{U}-1}%
=I_{0}.  \label{ee.12}
\end{equation}

We observe that equation (\ref{ee.12}) can be written in a linear form after
the change of the independent variable $d\kappa =\frac{e^{U}-1}{\left(
e^{U}\left( \rho ^{2}-1\right) +2\right) }d\rho $, and $U=U\left( \kappa
\right) .~$Thus, equation (\ref{ee.12}) becomes~$U_{\kappa }=I_{0}$ which is
nothing else than the conservation law for the maximal symmetric
second-order ordinary differential equation $U_{\kappa \kappa }=0$.

Last but not least we remark that find similar result if we apply first the
reduction of any one-dimensional Lie algebra of the three-dimensional
Abelian subalgebra and then we consider the $\left\{ X_{4}^{A}\right\} $.

\section{Lie symmetry analysis for equation $\mathcal{H}_{B}$}

\label{sec4}

In order to proceed with the analysis for equation $\mathcal{H}_{B}$ we
select the new set of independent variables $\left( z,\bar{z}\right) =\frac{1%
}{2}\left( u+v,i\left( u-v\right) \right) $, such that equation (\ref{ee.02}%
) to be written as follows%
\begin{equation}
u_{z\bar{z}}-\left( e^{u}\right) _{tt}=0.  \label{ee.14}
\end{equation}%
Application of the Lie symmetry conditions indicates that equation (\ref%
{ee.14}) admits the Lie symmetry vectors%
\begin{equation*}
X_{1}^{B}=\partial _{t}~,~X_{2}^{B}=t\partial _{t}+u\partial _{u},
\end{equation*}%
\begin{equation*}
X_{3}^{B}=\Phi \left( z\right) \partial _{z}-\Phi \left( z\right)
_{z}\partial _{u}~,~X_{4}^{B}=\Psi \left( \bar{z}\right) \partial _{z}-\Psi
\left( \bar{z}\right) _{\bar{z}}\partial _{u}.
\end{equation*}

The vector fields $X_{3}^{B}~,~X_{4}^{B}$ indicates the infinity number of
solutions for the Laplace operator $u_{z\bar{z}}$. The commutators and the
Adjoint representation for the admitted symmetry vectors are presented in
Tables \ref{hb1} and \ref{hb2} respectively. The Lie point symmetries, from
the finite Lie algebra $A_{2,1}$ plus the infinity algebra consisted by the
vector fields $X_{3}^{B}$ and $~X_{4}^{B}$. In Tables \ref{hb1} and \ref{hb2}
we assumed that functions $\Phi $ and $\Psi $ are specific and not
arbitrary. Because in general it holds $\left[ X_{3}^{B}\left( \Phi
_{1}\right) ,X_{3}^{B}\left( \Phi _{2}\right) \right] =X_{3}^{B}\left( \Phi
_{3}\right) $.

\begin{table}[tbp] \centering%
\caption{Commutator table for the Lie point symmetries of equation
(\ref{ee.02}).}%
\begin{tabular}{ccccc}
\hline\hline
$\left[ X_{I},X_{J}\right] $ & \textbf{$X^{B}$}$_{1}$ & \textbf{$X^{B}$}$%
_{2} $ & \textbf{$X^{B}$}$_{3}$ & \textbf{$X^{B}$}$_{4}$ \\ \hline
\textbf{$X^{B}$}$_{1}$ & $0$ & $X_{1}^{B}$ & $0$ & $0$ \\
\textbf{$X^{B}$}$_{2}$ & $-X_{1}^{B}$ & $0$ & $0$ & $0$ \\
\textbf{$X^{B}$}$_{3}$ & $0$ & $0$ & $0$ & $0$ \\
\textbf{$X^{B}$}$_{4}$ & $0$ & $0$ & $0$ & $0$ \\ \hline\hline
\end{tabular}%
\label{hb1}%
\end{table}%

\begin{table}[tbp] \centering%
\caption{Adjoint representation for the Lie point symmetries of equation
(\ref{ee.02}).}%
\begin{tabular}{ccccc}
\hline\hline
$Ad\left( e^{\left( \varepsilon \mathbf{X}_{i}\right) }\right) \mathbf{X}%
_{j} $ & \textbf{$X^{B}$}$_{1}$ & \textbf{$X^{B}$}$_{2}$ & \textbf{$X^{B}$}$%
_{3}$ & \textbf{$X^{B}$}$_{4}$ \\ \hline
\textbf{$X^{B}$}$_{1}$ & $X_{1}^{B}$ & $X_{2}^{B}-\varepsilon X_{1}^{B}$ & $%
X_{3}^{B}$ & $X_{4}^{B}$ \\
\textbf{$X^{B}$}$_{2}$ & $e^{\varepsilon }X_{1}^{B}$ & $X_{2}^{B}$ & $%
X_{3}^{B}$ & $X_{4}^{B}$ \\
\textbf{$X^{B}$}$_{3}$ & $X_{1}^{B}$ & $X_{2}^{B}$ & $X_{3}^{B}$ & $%
X_{4}^{B} $ \\
\textbf{$X^{B}$}$_{4}$ & $X_{1}^{B}$ & $X_{2}^{B}$ & $X_{3}^{B}$ & $%
X_{4}^{B} $ \\ \hline\hline
\end{tabular}%
\label{hb2}%
\end{table}%

Hence, from Tables \ref{hb1} and \ref{hb2} it follows that one-dimensional
optimal system is consisted by the following one-dimensional Lie algebras, $%
\left\{ X_{1}^{B}\right\} $,~$\left\{ X_{2}^{B}\right\} $,~$\left\{
X_{3}^{B}\right\} $, $\left\{ X_{4}^{B}\right\} $, $\left\{ X_{3}^{B}+\alpha
X_{4}^{B}\right\} $,~$\left\{ X_{1}^{B}+\alpha X_{3}^{B}\right\} $, $\left\{
X_{1}^{B}+\alpha X_{4}^{B}\right\} $, $\left\{ X_{2}^{B}+\alpha
X_{3}^{B}\right\} $, $\left\{ X_{2}^{B}+\alpha X_{4}^{B}\right\} $, $\left\{
X_{1}^{B}+\alpha X_{3}^{B}+\beta X_{4}^{B}\right\} $ and $\left\{
X_{2}^{B}+\alpha X_{3}^{B}+\beta X_{4}^{B}\right\} $. We proceed with the
reduction of the equation and the determination of similarity solutions.

Consider now reduction with the use of the symmetry vector $\left\{
X_{3}^{B}\right\} $, then it follows $u=-\ln \Phi \left( z\right) +\ln
U\left( t,\bar{z}\right) $ with reduced equation the
\begin{equation*}
U_{tt}=0~,~U\left( t,\bar{z}\right) =U_{1}\left( \bar{z}\right)
t+U_{0}\left( z\right) .
\end{equation*}%
We remark that the reduced equation is that of the free particle and it is
maximal symmetric, thus it admits eight Lie point symmetries which form the $%
sl\left( 3,R\right) $ Lie algebra. A similar result it follows if we assume
reduction with respect to the field $X_{4}^{B}$.

Let us assume now reduction with the field $\left\{
X_{1}^{B}+X_{4}^{B}\right\} $. The reduced equation is found to be%
\begin{equation}
U_{zT}+\left( e^{U}\right) _{TT}=0  \label{ee.15}
\end{equation}%
where $T=t-\int \frac{d\bar{z}}{\Psi \left( \bar{z}\right) }$ and $u=-\ln
\left( \Psi \left( \bar{z}\right) \right) +U\left( T,x\right) $.

Equation (\ref{ee.15}) admits the symmetry vectors $\bar{X}_{1}^{B}=\partial
_{T}~,~\bar{X}_{2}=T\partial _{T}+\partial _{U}$ and $\bar{X}_{3}^{B}=\Phi
\left( z\right) \partial _{z}-\Phi \left( z\right) _{z}\partial _{U}~$. \
Hence, application for the field $\bar{X}_{1}^{B}+\bar{X}_{3}^{B}$ gives the
similarity transformation $U\left( T,x\right) =-\ln \left( \Phi \left(
z\right) \right) +V\left( \tau \right) ~$, $\tau =T-\int \frac{dz}{\Phi
\left( z\right) }$, with reduced equation the maximal symmetric ordinary
differential equation%
\begin{equation*}
V_{\tau \tau }-\frac{e^{V}}{\left( 1-e^{V}\right) }\left( V_{\tau }\right)
^{2}=0\text{. }
\end{equation*}%
Moreover, application of the vector field $\bar{X}_{2}+\bar{X}_{3}^{B}$ in (%
\ref{ee.15}) provides the reduced equation%
\begin{equation}
V_{\lambda \lambda }\left( e^{V}-\lambda \right) +e^{V}\left( V_{\lambda
}\right) ^{2}-V_{\lambda }=0,  \label{ee.16}
\end{equation}%
where $\lambda =Te^{-S\left( z\right) }$,~$M\left( z\right) =\frac{1}{S_{,z}}
$, and $U\left( T,z\right) =S\left( z\right) +\ln \left( S\left( z\right)
_{z}\right) +V\left( Te^{-S\left( z\right) }\right) $.

Equation (\ref{ee.16}) is not maximal symmetric, however it can be
integrated and it can written in the equivalent form%
\begin{equation*}
\bar{V}_{\bar{\lambda}}=\frac{e^{\bar{\lambda}}\left( 1+\bar{V}\right) \bar{V%
}}{e^{\bar{\lambda}-1}}~,~\bar{\lambda}=V-\ln \lambda ~,~\bar{V}=\left(
V_{,\lambda }\lambda -1\right) ^{-1}.
\end{equation*}%
Therefore, if we do the change of variables$~d\lambda ^{\prime }=I_{0}\frac{%
e^{\bar{\lambda}}\left( 1+\bar{V}\right) \bar{V}}{e^{\bar{\lambda}-1}}d\bar{%
\lambda}$, the latter differential equation becomes $\bar{V}_{\lambda
^{\prime }}=I_{0}$, which is the conservation law for the maximal symmetric
differential equation$~\bar{V}_{\lambda ^{\prime }\lambda ^{\prime }}=0$.

\section{Lie symmetry analysis for equation $\mathcal{H}_{C}$}

\label{sec5}

As far as the Lie symmetries of $\mathcal{H}_{C}$ are concerned, they are
calculated%
\begin{equation*}
X_{1}^{C}=\partial _{t}~,~X_{2}^{C}=\partial _{y}~,~X_{3}^{C}=t\partial
_{t}+x\partial _{x}+y\partial _{y},
\end{equation*}%
\begin{equation*}
X_{4}^{C}\left( Z\left( y\right) \right) =Z\left( y\right) \left( \partial
_{x}-2\partial _{y}\right) +2Z\left( y\right) _{y}\partial _{u}.
\end{equation*}%
Hence, we can infer that equation $\mathcal{H}_{C}$ admits infinity Lie
symmetries. The nonzero commutators are~%
\begin{equation*}
\left[ X_{2}^{C},X_{3}^{C}\right] =X_{2}^{C}~,~~\left[ X_{2}^{C},X_{4}^{C}%
\left( Z\left( y\right) \right) \right] =X_{4}^{C}\left( Z\left( y\right)
_{y}\right) ~,
\end{equation*}%
\begin{equation*}
~\left[ X_{3}^{C},X_{4}^{C}\left( Z\left( y\right) \right) \right]
=X_{4}^{C}\left( yZ\left( y\right) _{y}-Z\left( y\right) \right) ~.
\end{equation*}%
$~$\ and
\begin{equation*}
\left[ X_{4}^{C}\left( Z\left( y\right) \right) ,X_{4}^{C}\left( W\left(
y\right) \right) \right] =2X_{4}^{C}\left( Y\left( y\right) \right) ~\text{%
with }Y\left( y\right) =W\left( y\right) Z\left( y\right) _{y}-Z\left(
y\right) W\left( y\right) _{y}.
\end{equation*}

We observe that for $Z\left( y\right) _{y}=0$, a finite dimensional Lie
algebra exist, the four-dimensional Lie algebra $A_{4,5}$ of equation $%
\mathcal{H}_{A}$.

Consider the application of the Lie point symmetries $\left\{
X_{1}^{C}+\alpha X_{2}^{C},X_{1}^{C}+\alpha \partial _{x}\right\} $, then,
the reduced equation is derived%
\begin{equation}
0=\left( \alpha \left( e^{U}-1\right) +2\beta +\alpha \beta e^{U}\right)
U_{\sigma \sigma }+\left( 1+\beta ^{2}\right) \alpha e^{U}\left( U_{\sigma
}\right) ^{2},  \label{ee.18}
\end{equation}%
where now $u=U\left( \sigma \right) $, $\sigma =y-\alpha t+\frac{\alpha }{%
\beta }x$. \ We observe that equation (\ref{ee.18}) is a maximal symmetric
second-order ordinary differential equation.

We proceed with the second reduction approach, where we apply the Lie
symmetries$~\left\{ X_{1}^{C}+\alpha X_{2}^{C},X_{1}^{C}+\alpha
X_{3}^{C}\right\} $. Hence, equation $\mathcal{H}_{C}$ is reduced to the
second-order ordinary differential equation%
\begin{equation*}
0=\left( \left( \alpha ^{2}+\omega ^{2}\right) e^{U}-\omega \left( \omega
+2\right) \right) U_{\omega \omega }+\left( \alpha ^{2}+\omega ^{2}\right)
e^{U}\left( U_{\omega }\right) ^{2}+2\left( \omega \left( e^{U}-1\right)
-1\right) \left( U_{\omega }\right) ,
\end{equation*}%
where $u=U\left( \omega \right) $ and $\omega =\frac{y+\alpha t}{x}$. The
latter equation can be integrated as follows%
\begin{equation}
U_{\omega }\left( \left( \alpha ^{2}+\omega ^{2}\right) e^{U}-\omega \left(
\omega +2\right) \right) =I_{0}
\end{equation}%
which can be written in the equivalent form$~U_{\varpi }=I_{0}$.

Let us assume reduction with respect to the Lie symmetry vector $\left\{
X_{1}^{C}+X_{4}^{C}\left( Z\left( y\right) \right) \right\} $. The
similarity transformation is $u=U\left( T,X\right) $ with $T=t-\int \frac{%
d\chi }{Z\left( y+2x-2\chi \right) }$ and $X=y+2x$, while the reduced
equation is
\begin{equation}
\left( U_{TT}+\left( U_{T}\right) ^{2}+4\left( U_{YY}+\left( U_{Y}\right)
^{2}\right) \right) e^{U}+2U_{TY}=0.  \label{ee.20}
\end{equation}%
The latter equation admits the reduced symmetry vectors $\left\{ \partial
_{T},\partial _{Y},T\partial _{T}+Y\partial _{Y}\right\} $. \ It follows
that reduction with respect to the symmetry vector $\left\{ \partial
_{T}+\beta \partial _{Y}\right\} $ gives $U=V\left( z\right) ,~z=\frac{1}{%
\beta }T-Y$ in which $V\left( z\right) $ is a solution of the maximal
symmetric second-order ordinary differential equation%
\begin{equation*}
\left( \left( 4+\beta ^{2}\right) e^{V}-2\beta \right) V_{zz}+\left( \beta
^{2}+4\right) e^{V}\left( V_{z}\right) ^{2}=0.
\end{equation*}%
On the other hand, reduction of equation (\ref{ee.20}) with respect to the
similarity transformation provided by $T\partial _{T}+Y\partial _{Y}$ gives%
\begin{equation*}
\left( \left( 4\lambda ^{2}+1\right) e^{V}-2\lambda \right) V_{\lambda
\lambda }+\left( 4\lambda ^{2}+1\right) e^{V}\left( V_{\lambda }\right)
^{2}+\left( 8\lambda e^{V}-2\right) V_{\lambda }=0,
\end{equation*}%
that is%
\begin{equation*}
V_{\lambda }\left( \left( 4\lambda ^{2}+1\right) e^{V}-2\lambda \right)
=I_{0},
\end{equation*}%
which can be written as a maximal symmetric second-order differential
equation.

\section{Lie symmetry analysis for equation $\mathcal{H}_{D}$}

\label{sec6}

We proceed our analysis with the derivation of the Lie symmetry vectors for
equation $\mathcal{H}_{D}$. \ The application of the Lie symmetry condition
provides that equation $\mathcal{H}_{D}$ admits infinity number of Lie
symmetries as they described by the following families of vector fields%
\begin{eqnarray*}
X_{1}^{D}\left( \Phi \left( t\right) \right) &=&\Phi \left( t\right)
\partial _{t}+\left( \frac{1}{6}\Phi _{tt}y^{2}+\frac{1}{3}x\Phi \right)
\partial _{x}+\frac{2}{3}y\Phi _{t}\partial _{y} \\
&&-\frac{1}{3}\left( \Phi _{tt}x+\frac{1}{2}\Phi _{ttt}y^{2}-2\Phi
_{t}u\right) \partial _{u}~,~
\end{eqnarray*}%
\begin{equation*}
X_{2}^{D}\left( \Psi \left( t\right) \right) =\Psi \left( t\right) \partial
_{x}-\Psi \left( t\right) _{t}\partial _{u}~,~
\end{equation*}%
\begin{equation*}
X_{3}^{D}\left( \Sigma \left( t\right) \right) =\frac{1}{2}\Sigma
_{t}y\partial _{x}+\Sigma \partial _{y}-\frac{1}{2}\Sigma _{tt}y\partial
_{u},
\end{equation*}%
and%
\begin{equation*}
X_{4}^{D}=2x\partial _{x}+\partial _{y}+2u\partial _{u}.
\end{equation*}

The nonzero commutators of the admitted Lie symmetries are%
\begin{equation*}
\left[ X_{1}^{D}\left( \Phi \right) ,X_{2}^{D}\left( \Psi \right) \right]
=X_{2}^{D}\left( \Phi \Psi _{t}-\frac{1}{3}\Phi _{t}\Psi \right) ,
\end{equation*}%
\begin{equation*}
\left[ X_{1}^{D}\left( \Phi \right) ,X_{3}^{D}\left( \Sigma \right) \right]
=X_{3}^{D}\left( \frac{2}{3}\Sigma \Phi _{t}-3\Phi \Sigma _{t}\right) ,
\end{equation*}%
\begin{equation*}
\left[ X_{2}^{D}\left( \Psi \right) ,X_{4}^{D}\right] =2X_{2}^{D}\left( \Psi
\right) ~,~\left[ X_{3}^{D}\left( \Sigma \right) ,X_{4}^{D}\right]
=X_{3}^{D}\left( \Sigma \right) ,
\end{equation*}%
\begin{equation*}
\left[ X_{1}^{D}\left( \Phi \left( t\right) \right) ,X_{1}^{D}\left( \bar{%
\Phi}\left( t\right) \right) \right] =X_{1}^{D}\left( \Phi \bar{\Phi}_{t}-%
\bar{\Phi}\Phi _{t}\right) ,
\end{equation*}%
\begin{equation*}
\left[ X_{3}^{D}\left( \Sigma \left( t\right) \right) ,X_{3}^{D}\left( \bar{%
\Sigma}\left( t\right) \right) \right] =X_{2}^{D}\left( \bar{\Sigma}\Sigma
_{t}-\Sigma \Sigma _{t}\right) .
\end{equation*}

For $\Phi \left( t\right) =1~,~\Psi \left( t\right) =1$ and $\Sigma \left(
t\right) =1$, we find the four-dimensional subalgebra%
\begin{equation*}
\bar{X}_{1}^{D}=\partial _{t}~,~\bar{X}_{2}^{D}=\partial _{x}~,~\bar{X}%
_{3}^{D}=\partial _{y}\text{~},~X_{4}^{D},
\end{equation*}%
which form the $A_{3,3}\otimes A_{1}$ Lie algebra. The commutators and the
Adjoint representation of the four dimensional Lie algebra $\left\{ \bar{X}%
_{1}^{D},\bar{X}_{2}^{D},\bar{X}_{3}^{D},X_{4}^{D}\right\} $ are presented
in Tables \ref{hd1} and \ref{hd2}.

\begin{table}[tbp] \centering%
\caption{Commutators for the elements which form the finite Lie algebra of equation
(\ref{ee.04}).}%
\begin{tabular}{ccccc}
\hline\hline
$\left[ X_{I},X_{J}\right] $ & \textbf{$\bar{X}^{D}$}$_{1}$ & \textbf{$\bar{X%
}^{D}$}$_{2}$ & \textbf{$\bar{X}^{D}$}$_{3}$ & \textbf{$X^{D}$}$_{4}$ \\
\hline
\textbf{$\bar{X}^{D}$}$_{1}$ & $0$ & $0$ & $0$ & $0$ \\
\textbf{$X^{D}$}$_{2}$ & $0$ & $0$ & $0$ & $2\bar{X}_{2}^{D}$ \\
\textbf{$\bar{X}^{D}$}$_{3}$ & $0$ & $0$ & $0$ & $\bar{X}_{3}^{D}$ \\
$\mathbf{X}$\textbf{$^{D}$}$_{4}$ & $0$ & $-2\bar{X}_{2}^{D}$ & $-\bar{X}%
_{3}^{D}$ & $0$ \\ \hline\hline
\end{tabular}%
\label{hd1}%
\end{table}%

\begin{table}[tbp] \centering%
\caption{Adjoint representation  for the elements which form the finite Lie algebra
(\ref{ee.04}).}%
\begin{tabular}{ccccc}
\hline\hline
$Ad\left( e^{\left( \varepsilon \mathbf{X}_{i}\right) }\right) \mathbf{X}%
_{j} $ & \textbf{$\bar{X}^{D}$}$_{1}$ & \textbf{$\bar{X}^{D}$}$_{2}$ &
\textbf{$\bar{X}^{D}$}$_{3}$ & \textbf{$X^{D}$}$_{4}$ \\ \hline
\textbf{$\bar{X}^{D}$}$_{1}$ & $\bar{X}_{1}^{D}$ & $\bar{X}_{2}^{D}$ & $\bar{%
X}_{3}^{D}$ & $X_{4}^{D}$ \\
\textbf{$\bar{X}^{D}$}$_{2}$ & $\bar{X}_{1}^{D}$ & $\bar{X}_{2}^{D}$ & $\bar{%
X}_{3}^{D}$ & $X_{4}^{D}-2\varepsilon \bar{X}_{2}^{D}$ \\
\textbf{$\bar{X}^{D}$}$_{3}$ & $\bar{X}_{1}^{D}$ & $\bar{X}_{2}^{D}$ & $\bar{%
X}_{3}^{D}$ & $X_{4}^{D}-\varepsilon \bar{X}_{3}^{D}$ \\
\textbf{$X^{D}$}$_{4}$ & $\bar{X}_{1}^{D}$ & $e^{2\varepsilon }\bar{X}%
_{2}^{D}$ & $e^{\varepsilon }\bar{X}_{3}^{D}$ & $X_{4}^{D}$ \\ \hline\hline
\end{tabular}%
\label{hd2}%
\end{table}%

We proceed with the application of the Lie symmetries for the finite Lie
algebra $A_{3,3}\otimes A_{1}$ for the reduction of the partial differential
equation~$\mathcal{H}_{D}$. From table \ref{hd2} we derive the
one-dimensional optimal system, which is consisted by the one-dimensional
Lie algebras $\left\{ \bar{X}_{1}^{D}\right\} $, $\left\{ \bar{X}%
_{2}^{D}\right\} $, $\left\{ \bar{X}_{3}^{D}\right\} $, $\left\{ \bar{X}%
_{4}^{D}\right\} $, $\left\{ \bar{X}_{1}^{D}+\alpha \bar{X}_{2}^{D}\right\} $%
, $\left\{ \bar{X}_{1}^{D}+\alpha \bar{X}_{3}^{D}\right\} $, $\left\{ \bar{X}%
_{1}^{D}+\alpha \bar{X}_{4}^{D}\right\} $, $\left\{ \bar{X}_{2}^{D}+\alpha
\bar{X}_{3}^{D}\right\} $, $\left\{ \bar{X}_{1}^{D}+\alpha \bar{X}%
_{2}^{D}+\beta \bar{X}_{3}^{D}\right\} $.

Therefore, by applying the Lie symmetries $\left\{ \bar{X}_{1}^{D}+\alpha
\bar{X}_{2}^{D}\right\} $ , $\left\{ \bar{X}_{1}^{D}+\beta \bar{X}%
_{3}^{D}\right\} $ for the reduction of equation (\ref{ee.04}) we end with
the second-order ordinary differential equation%
\begin{equation}
\left( \alpha ^{2}+\beta ^{2}U+\alpha \beta ^{2}\right) U_{\sigma \sigma
}+\beta ^{2}\left( U_{\sigma }\right) ^{2}=0,  \label{ee.21}
\end{equation}%
with $u=U\left( \sigma \right) $ and $\sigma =y-\beta t+\frac{\beta }{\alpha
}x$. Equation (\ref{ee.21}) is maximal symmetric and can be linearized.

On the other hand, from the Lie symmetry $\left\{ \bar{X}_{2}^{D}+\bar{X}%
_{3}^{D}\right\} $ we find the second-order partial differential equation
\begin{equation}
U_{Yt}+U_{YY}+UU_{YY}+\left( U_{Y}\right) ^{2}=0.  \label{ee.22}
\end{equation}%
where $u=U\left( t,Y\right) $, $Y=y-x$. Equation (\ref{ee.22}) admits
infinity Lie symmetries consisted by the vector fields $Y_{1}^{D}=\partial
_{t}$~,~$Y_{2}^{D}=t\partial _{t}+\left( U-1\right) \partial _{U}$,~$%
Y_{3}^{D}=Y\partial _{Y}+\left( U+1\right) \partial _{U}$ , $%
Y_{4}^{D}=t^{2}+tY\partial _{t}+\left( Y-t\left( U+1\right) \right) \partial
_{U}$ and $Y_{5}^{D}=\zeta \left( t\right) \partial _{Y}+\zeta \left(
t\right) _{t}\partial _{U}$. Vector fields $\left\{
Y_{1}^{D},Y_{2}^{D},Y_{4}^{D}\right\} $ form the $sl\left( 2,R\right) $ Lie
algebra.

Hence, from the vector field $\left\{ Y_{1}^{D}+\alpha Y_{3}^{D}\right\} $
it follow the similarity transformation $U=-1+V\left( \kappa \right)
e^{\alpha t}$,~$\kappa =Ye^{-\alpha t}$, with reduced equation the
differential equation $V_{\kappa \kappa }+\frac{1}{V-\kappa \alpha }\left(
V_{\kappa }\right) ^{2}=0$, which can be integrated further~%
\begin{equation}
N_{\lambda }=\frac{N}{\lambda }\left( 1+\alpha N\right) ^{2~}\text{with}%
~N=\left( V_{\kappa }-\alpha \right) ^{-1}~\text{and}~\kappa =V-\kappa
\alpha \text{.}  \label{ee.23}
\end{equation}%
Easily, equation (\ref{ee.23}) can be written as $N_{\bar{\lambda}\bar{%
\lambda}}=0$. A similar result follows, if we perform the reduction of
equation (\ref{ee.22}) with the rest of the symmetry vectors.

\section{Lie symmetry analysis for equation $\mathcal{H}_{E}$}

\label{sec7}

The fifth equation of our study, namely equation $\mathcal{H}_{E}$ admits
the following Lie symmetries%
\begin{equation*}
X_{1}^{E}\left( \Phi \left( t\right) \right) =\Phi \left( t\right) \partial
_{t}+\frac{1}{2}y\Phi _{t}\partial _{y}+\left( \frac{1}{4}\Phi _{tt}y-\frac{1%
}{2}\Phi _{t}u\right) \partial _{u},
\end{equation*}%
\begin{equation*}
X_{2}^{E}=\partial _{x}~,~X_{3}^{E}\left( \Sigma \left( t\right) \right)
=\Sigma \left( t\right) \partial _{y}+\frac{1}{2}\Sigma _{t}\partial _{u},~
\end{equation*}%
\begin{equation*}
X_{4}^{E}=2x\partial _{x}+y\partial _{y}+u\partial _{u}\text{.}
\end{equation*}%
The Lie symmetries form an infinity Lie algebra, with nonzero commutators%
\begin{equation*}
\left[ X_{1}^{E}\left( \Phi \left( t\right) \right) ,X_{3}^{E}\left( \Sigma
\left( t\right) \right) \right] =X_{3}^{D}\left( \Sigma \Phi _{t}-2\Phi
\Sigma _{t}\right) ~,~\left[ X_{2}^{E},X_{4}^{E}\right] =2X_{2}^{E}~,
\end{equation*}%
\begin{equation*}
\left[ X_{3}^{E}\left( \Sigma \left( t\right) \right) ,X_{4}^{E}\right]
=X_{3}^{E}\left( \Sigma \left( t\right) \right) ,
\end{equation*}%
and%
\begin{equation*}
\left[ X_{1}^{E}\left( \Phi \left( t\right) \right) ,X_{1}^{E}\left( \bar{%
\Phi}\left( t\right) \right) \right] =X_{1}^{E}\left( \Phi \bar{\Phi}_{t}-%
\bar{\Phi}\Phi _{t}\right) .
\end{equation*}%
The four-dimensional finite algebra $A_{3,3}\otimes A_{1}$ follows for $\Phi
\left( t\right) =1~$and$~\Sigma \left( t\right) =1$, that is, $%
A_{3,3}\otimes A_{1}$ is consisted by the Lie symmetry vectors $\left\{ \bar{%
X}_{1}^{E},X_{2}^{E},\bar{X}_{3}^{E},X_{4}^{E}\right\} $. We proceed with
the application of the Lie symmetry vectors and the determination of the
similarity transformations

From the symmetry vectors $\left\{ \bar{X}_{1}^{E}+\alpha X_{2}^{E}\right\} $%
,~$\left\{ \bar{X}_{1}^{E}+\beta \bar{X}_{3}^{E}\right\} $ we find the
similarity transformation $u=U\left( \sigma \right) $,~$\sigma =y-\beta t+%
\frac{\beta }{\alpha }x$ with reduced equation the maximal symmetric equation%
\begin{equation*}
\left( 2U\beta +\alpha -\beta ^{2}\right) U_{\sigma \sigma }+2\beta \left(
U_{\sigma }\right) ^{2}=0.
\end{equation*}

On the other hand, reduction with the symmetry vector $\left\{ X_{2}^{E}+%
\bar{X}_{3}^{E}\right\} $ provides the second-order partial differential
equation%
\begin{equation}
V_{Yt}-V_{YY}+2VV_{YY}+2\left( V_{Y}\right) ^{2}=0,  \label{ee.24}
\end{equation}%
where $u=V\left( t,Y\right) $, $Y=y-x$. Equation (\ref{ee.24}) is of the
form of equation (\ref{ee.22}).\ Indeed if we replace in (\ref{ee.24}) $V=-%
\frac{1}{2}U$ and $t\rightarrow -t$ equation (\ref{ee.24}) is written in the
form of equation (\ref{ee.22}).

\section{Conclusions}

\label{sec8}

We applied the Lie symmetry analysis for a family of five partial
differential equations of the form (\ref{ee.00}) which are integrable with
the method of hydrodynamic reductions. In particular, we determined the Lie
point symmetries and we studied the algebraic properties of the admitted
symmetries. Moreover, from the invariant functions provided by the Lie
symmetries we defined similarity transformations which were used to reduce
the number of independent variables for the differential equations. With the
application of two different similarity transformations we were able to
reduce the partial differential equations into a second-order ordinary
differential equation. We summarize this result in the following proposition.

\textbf{Proposition 1:} The five partial differential equations $\mathcal{H}%
_{A}$, $\mathcal{H}_{B}$, $\mathcal{H}_{C}$, $\mathcal{H}_{D}$ and $\mathcal{%
H}_{E}$ which are integrable with the method of hydrodynamic reductions can
be linearized with the use of similaritry transformations given by the Lie
point symmetries.

Equation $\mathcal{H}_{A}$ admits a finite Lie algebra of dimension four,
while the rest differential equations, $\mathcal{H}_{B}$, $\mathcal{H}_{C}$,
$\mathcal{H}_{D}$ and $\mathcal{H}_{E}$ admit infinity Lie point symmetries
which however are constructed by four generic vector fields. The application
of the Lie point symmetries for these equations indicates that these five
equations posses a common feature, they are reduced to a maximal symmetric
ordinary differential equation which can be linearized. We show that this is
possible not only when we investigate for \textquotedblleft
travel-wave\textquotedblright\ solutions but also for more general
reductions.

In a future study we plan to investigate by applying the theory of Lie
symmetries and other differential equations which are integrable by the
method of hydrodynamic reductions.

\bigskip

\textbf{Data availability statement}

\textit{Data sharing not applicable to this article as no datasets were
generated or analysed during the current study.}

\end{document}